\documentclass[aps,prl,reprint,twocolumn,superscriptaddress,floatfix,preprintnumbers]{revtex4-2}
\usepackage{graphicx,amsmath,amsfonts,amssymb,amsthm,xr,physics}
\usepackage{color,dsfont,upgreek}
\usepackage{mathrsfs}
\usepackage{mathtools}
\usepackage{bbold}
\usepackage{slashed}
\usepackage{float}
\usepackage[caption=false]{subfig}
\usepackage[dvipsnames]{xcolor}
\usepackage[bookmarks=true,colorlinks,linkcolor=OrangeRed,urlcolor=NavyBlue,citecolor=RoyalBlue]{hyperref}
\usepackage[capitalize]{cleveref}
\usepackage{orcidlink}

\newcounter{supplem}

\usepackage{graphicx}
\usepackage{dcolumn}
\usepackage{bm}

\usepackage{lineno}

\newcommand{\ulm}{Institute for Complex Quantum Systems, Ulm University, 89069 Ulm, Germany}
\newcommand{\iqst}{Center for Integrated Quantum Science and Technology (IQST), Ulm-Stuttgart, Germany}
\newcommand{\mitus}{Center for Theoretical Physics - a Leinweber Institute, Massachusetts Institute of Technology, Cambridge, MA 02139, USA}

\begin{document}

\title{Quantum Phase Diagram of the $2+1$D Untruncated SU$(2)$ Lattice Gauge Theory with Dynamical Fermions}

\author{Gabriel Rouxinol$^{\orcidlink{0009-0004-8147-9814}}$}
\affiliation{Department of Physics and Arnold Sommerfeld Center for Theoretical Physics (ASC), Ludwig Maximilian University of Munich, 80333 Munich, Germany}
\affiliation{Munich Center for Quantum Science and Technology (MCQST), 80799 Munich, Germany}

\author{Julian Bender${}^{\orcidlink{0000-0003-4920-7849}}$}
\affiliation{\mitus}

\author{Patrick Emonts${}^{\orcidlink{0000-0002-7274-4071}}$}
\affiliation{\ulm}
\affiliation{\iqst}

\author{Michele Grossi$^{\orcidlink{0000-0003-1718-1314}}$}
\affiliation{European Organisation for Nuclear Research (CERN), 1211 Geneva, Switzerland}

\author{Jad C.~Halimeh${}^{\orcidlink{0000-0002-0659-7990}}$}
\email{jad.halimeh@lmu.de}
\affiliation{Department of Physics and Arnold Sommerfeld Center for Theoretical Physics (ASC), Ludwig Maximilian University of Munich, 80333 Munich, Germany}
\affiliation{Max Planck Institute of Quantum Optics, 85748 Garching, Germany}
\affiliation{Munich Center for Quantum Science and Technology (MCQST), 80799 Munich, Germany}
\affiliation{Department of Physics, College of Science and Technology, Kyung Hee University, Seoul 02447, Republic of Korea}

\date{\today}

\begin{abstract}
Non-Abelian gauge theories with dynamical matter govern the strong interaction and a broad class of strongly correlated quantum systems, yet their ground-state properties remain difficult to obtain from first principles. 
Using a continuous-group variational Monte Carlo approach that retains the full SU$(2)$ gauge field without truncation, we determine the ground-state behavior of the SU$(2)$ lattice gauge theory with staggered fermions on an $L\times L$ square lattice. 
Treating the magnetic and electric couplings $\lambda$ and $g^2$ independently, we find a magnetic-flux transition at $\lambda^\ast=-0.040\pm 0.005$, with no resolvable drift of the transition point as the electric coupling is varied. 
Along the physical coupling line $\lambda=4/g^2$, for $L=4,6,8$, we uncover a gauge--matter delocalization crossover from a flux-disordered regime at strong electric coupling to an ordered unity-flux regime at weak coupling. 
The chiral condensate, a gauge-invariant Wilson-line meson correlator, and the local color density consistently reveal the emergence of coherent gauge-assisted matter dynamics. 
Together, these results provide a unified physical picture of how magnetic-flux ordering and fermionic coherence develop in an untruncated non-Abelian lattice gauge theory.
\end{abstract}

\maketitle

\textbf{\textit{Introduction.---}}
Gauge theories underlie the Standard Model of particle physics, describing the interactions of elementary particles mediated by gauge bosons~\cite{Weinberg1995QuantumTheoryFields,peskin2018introduction,srednicki2007quantum}. Their lattice formulation provides the standard nonperturbative framework for studying strongly interacting quantum field theories, including confinement in high-energy physics~\cite{wilsonConfinementQuarks1974,grossAsymptoticallyFreeGauge1973}. Beyond particle physics, lattice gauge theories (LGTs) also emerge as effective descriptions of strongly correlated quantum matter~\cite{wen2004quantum,Balents2010SpinLiquidsFrustrated,Savary2016QuantumSpinLiquids,Fradkin2013FieldTheoriesCondensed,affleckLargenLimitHeisenbergHubbard1988,wenMeanfieldTheorySpinliquid1991,leeDopingMottInsulator2006} and provide paradigmatic settings for exotic nonequilibrium quantum many-body phenomena~\cite{Smith2017AbsenceOfErgodicity,Brenes2018ManyBodyLocalization,Budde2024QuantumManyBodyScars,osborneQuantumManyBodyScarring2024,Cataldi2025DisorderFreeLocalizationFragmentation-1}.

\begin{figure}[t!]
    \centering
    \includegraphics[width=1.0\linewidth]{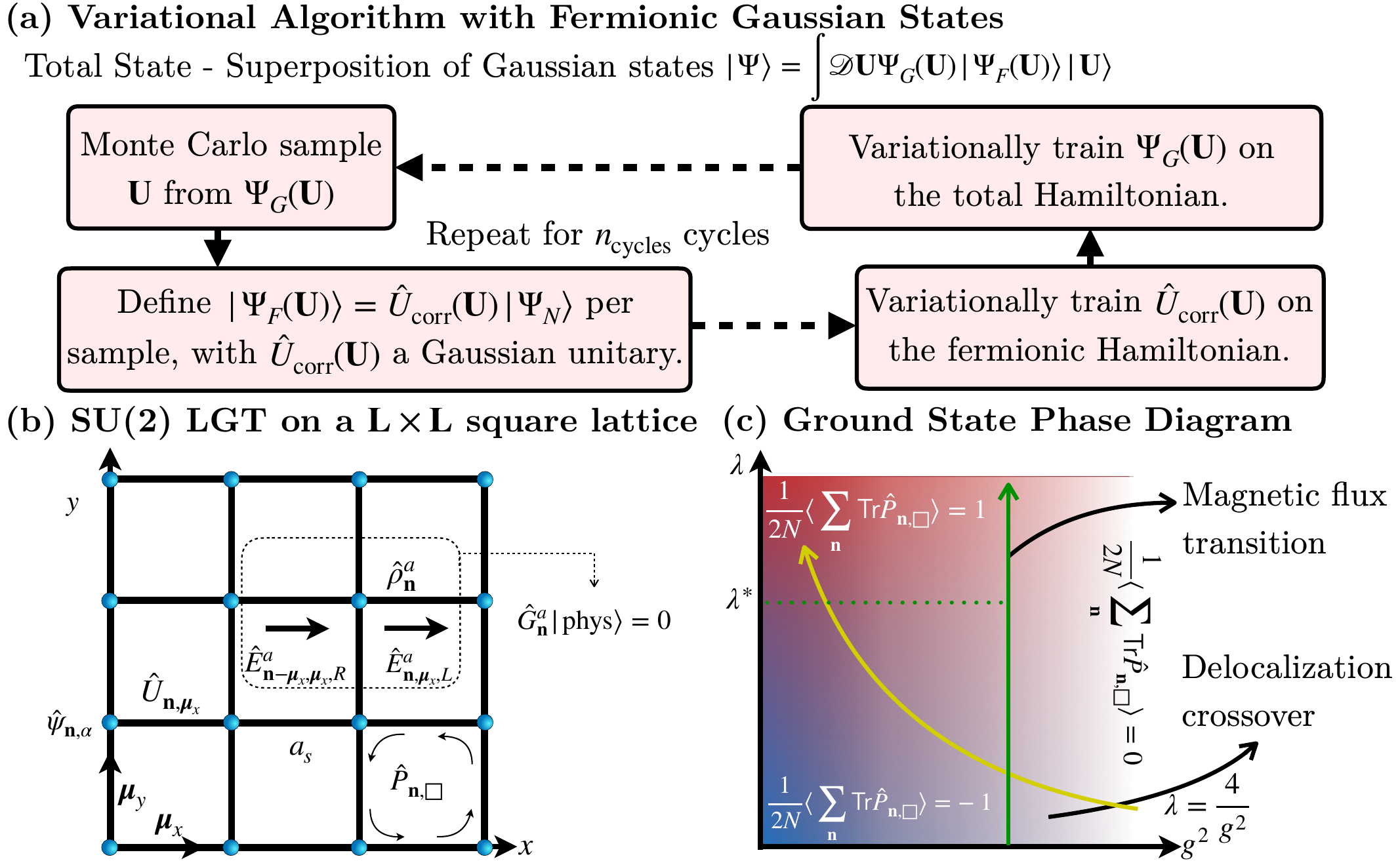}
    \caption{Variational framework, model, and principal physical results. (a) Gauge configurations $\mathbf U$ are sampled from $\Psi_G(\mathbf U)$, each defining a gauge-covariant Gaussian fermionic state, and $\hat U_{\mathrm{corr}}$ and $\Psi_G$ are optimized using the fermionic and full Hamiltonians, respectively. (b) The $2+1$D square lattice. (c) Ground-state behavior in the $(\lambda,g^2)$ plane, showing the magnetic-flux transition (green) and gauge--matter delocalization crossover (yellow) discussed in the text.}
    \label{fig:scheme}
\end{figure}

Among these models, non-Abelian LGTs with dynamical matter constitute the minimal setting in which confinement and correlated gauge--matter physics can be investigated beyond Abelian theories~\cite{banerjeeAtomicQuantumSimulation2013,tagliacozzoSimulationNonAbelianGauge2013,cataldiRealTimeStringDynamics2025}. In $2+1$ dimensions, SU$(2)$ is the simplest continuous non-Abelian gauge theory while already capturing qualitative features relevant to QCD~\cite{atasSU2HadronsQuantum2021,tepermathrmSUNGaugeTheories1998,karabaliGaugeinvariantHamiltonianAnalysis1996,feynmanQualitativeBehaviorYangMills1981}. However, obtaining its ground-state properties from first principles remains challenging. Euclidean Monte Carlo methods are hindered in important fermionic regimes by the sign problem, particularly at finite chemical potential or in the presence of a topological $\theta$ term~\cite{fodorNewMethodStudy2002,troyerComputationalComplexityFundamental2005,vicariThthDependenceSUNSUN2009,philipsenLatticeQCDFinite2007}, motivating Hamiltonian approaches~\cite{kogutHamiltonianFormulationWilsons1975,banulsSimulatingLatticeGauge2020}. Even within the Hamiltonian formulation, simultaneously enforcing gauge invariance, incorporating dynamical matter, and retaining the full continuous gauge group without truncation remains a computational challenge. The same difficulty limits complementary approaches based on quantum simulation~\cite{byrnesSimulatingLatticeGauge2006, dalmonteLatticeGaugeTheory2016, zoharQuantumSimulationsLattice2015, aidelsburgerColdAtomsMeet2021a, zoharQuantumSimulationLattice2021,
barataMediumInducedJet2022,klcoStandardModelPhysics2022,barataQuantumSimulationInmedium2023,barataRealtimeDynamicsHyperon2024,bauerQuantumSimulationHighEnergy2023, Bauer2023QuantumSimulationFundamental,
dimeglioQuantumComputingHighEnergy2024, Cheng2024EmergentGaugeTheory, Cohen2021QuantumAlgorithmsTransport,Barata2025ProbingCelestialEnergy, leeQuantumComputingEnergy2025, turroClassicalQuantumComputing2024,Halimeh2023ColdatomQuantumSimulators,bauerEfficientUseQuantum2025,halimehQuantumSimulationOutofequilibrium2025} and tensor-network methods~\cite{ricoTensorNetworksLattice2014,banulsEfficientBasisFormulation2017,magnificoTensorNetworksLattice2025,emontsFindingGroundState2023,orusPracticalIntroductionTensor2014,cataldiHamiltonianLatticeGauge2025,felserTwoDimensionalQuantumLinkLattice2020,cataldiSimulating$2+1mathrmD$SU22024}, particularly in higher dimensions~\cite{gyawaliObservationDisorderfreeLocalization2025,Cochran2025VisualizingDynamicsCharges,gonzalez-cuadraObservationStringBreaking2025,crippaAnalysisConfinementString2024,cobosRealTimeDynamics2025,xuStringBreakingDynamics2025,joshiObservationGenuine$2+1$D2026} and for untruncated continuous gauge groups~\cite{Martinez2016RealtimeDynamicsLattice}. 

Machine-learning variational methods have recently emerged as a powerful framework for quantum many-body problems~\cite{carleoSolvingQuantumManybody2017,mcmillanGroundStateLiquid1965,arisueVariationalStudyVacuum1983,chinExactGroundstateProperties1985,medvidovicNeuralnetworkQuantumStates2024,pfauAccurateComputationQuantum2024,wuRealNeuralNetwork2023,wuNNQSTransformerEfficientScalable2023,liImprovedOptimizationNeuralnetwork2024,wangVariationalOptimizationAmplitude2024,nysRealtimeQuantumDynamics2024,louNeuralWaveFunctions2024,luVariationalNeuralTensor2025,wuDeepQuarkDeepNeuralNetworkApproach2026}. For LGTs, magnetic-basis variational Monte Carlo has enabled accurate ground-state calculations of pure-gauge SU$(2)$ theories in both $2+1$D and $3+1$D without truncating the gauge group~\cite{spriggsAccurateGroundStates2026}. Independently, Gaussian variational and continuous-group Monte Carlo methods have incorporated dynamical matter in Abelian or lower-dimensional settings~\cite{salaVariationalStudyU12018,benderVariationalMonteCarlo2023}, while gauge-equivariant neural-network and matrix-model approaches provide alternative variational formulations of non-Abelian theories~\cite{luoGaugeEquivariantNeural2021,bodendorferVariationalMonteCarlo2025,favoniApplicationsLatticeGauge2022,luoGaugeinvariantAnyonicsymmetricAutoregressive2023,rayatGraphNeuralNetworks2026}. Here we combine these developments into a variational Monte Carlo framework for the $2+1$D SU$(2)$ lattice gauge theory with staggered fermions, in which a neural-network gauge wave function and a gauge-covariant Gaussian fermionic ansatz are optimized jointly in the magnetic basis. Figure~\ref{fig:scheme} summarizes both the variational framework and the principal physical results, while the algorithmic construction and benchmarks are presented in the companion paper~\cite{rouxinolGaugeEquivariantGaussian2026}.

In this Letter, we use this framework to investigate how the interplay between magnetic-flux ordering and dynamical matter shapes the ground-state physics of the $2+1$D SU$(2)$ lattice gauge theory. Treating the magnetic coupling $\lambda$ and electric coupling $g^2$ as independent parameters isolates the competition between magnetic and fermionic energy scales before restoring the physical relation $\lambda=4/g^2$. Within this extended parameter space, we identify a magnetic-flux transition at $\lambda^\ast=-0.040\pm0.005$, with no resolved dependence of $\lambda^\ast$ on $g^2$ within our numerical resolution. Along the physical coupling line, we uncover a crossover from a flux-disordered regime at strong electric coupling to an ordered unity-flux regime at weak coupling, accompanied by a gauge--matter delocalization crossover. 
The chiral condensate, a gauge-invariant Wilson-line meson correlator,
and the local color density consistently reveal how the reduced energetic
penalty for electric-field fluctuations promotes coherent gauge-assisted
matter dynamics, driving the fermionic state away from the N\'eel
reference and toward the free-fermion limit obtained at
$\hat U_{\mathbf n,\boldsymbol{\mu}_k}=\mathbb I$.

\textbf{\textit{Model.---}}
We study the $2+1$-dimensional SU$(2)$ LGT coupled to dynamical fermions on an $L\times L$ square lattice with periodic boundary conditions. In the Hamiltonian formulation with staggered fermions~\cite{kogutHamiltonianFormulationWilsons1975},
\begin{align}
\label{eqn:HamiltonianLGT}
    \hat{H} &= \hat{H}_E+\hat{H}_B+\hat{H}_m+\hat{H}_t,\\
\nonumber
    \hat{H}_E &= \frac{g^2}{2a_s}\sum_{\mathbf{n},k,a}
    \hat{E}_{\mathbf{n},\boldsymbol{\mu}_k}^a
    \hat{E}_{\mathbf{n},\boldsymbol{\mu}_k}^a,\\
\nonumber
    \hat{H}_B &= \frac{\lambda}{a_s}\sum_{\mathbf{n}}
    \left(1-\frac{1}{2}\Tr\hat{P}_{\mathbf{n},\Box}\right),\\
\nonumber
    \hat{H}_m &= m\sum_{\mathbf{n},\alpha}
    (-1)^{n_x+n_y}
    \hat{\psi}_{\mathbf{n},\alpha}^{\dagger}
    \hat{\psi}_{\mathbf{n},\alpha},\\
\nonumber
    \hat{H}_t &= -\frac{it}{2a_s}
    \sum_{\mathbf{n},k,\alpha,\beta}
    \left(
    \hat{\psi}_{\mathbf{n},\alpha}^{\dagger}
    \hat{U}_{\mathbf{n},\boldsymbol{\mu}_k}^{\alpha\beta}
    \hat{\psi}_{\mathbf{n}+\boldsymbol{\mu}_k,\beta}
    \eta_{\mathbf{n},\boldsymbol{\mu}_k}
    -\text{H.c.}
    \right).
\end{align}
Here $\mathbf{n}=(n_x,n_y)$ labels the $N=L^2$ sites, $\boldsymbol{\mu}_k$ is the unit vector in direction $k\in(x,y)$, and $\hat{U}_{\mathbf{n},\boldsymbol{\mu}_k}$ is the SU$(2)$ gauge operator on link $(\mathbf{n},\boldsymbol{\mu}_k)$. The set of all links is denoted by $\mathbf U$. The staggered fermion operators $\hat{\psi}_{\mathbf{n},\alpha}$ carry a fundamental-representation color index $\alpha$ and satisfy
$\{\hat{\psi}_{\mathbf{n},\alpha},
\hat{\psi}^{\dagger}_{\mathbf{n}^{\prime},\beta}\}
=\delta_{\mathbf{n},\mathbf{n}^{\prime}}\delta_{\alpha,\beta}$.
The staggering factor is
$\eta_{\mathbf{n},\boldsymbol{\mu}_k}=(-1)^{n_x}$ for $k=y$
and $\eta_{\mathbf{n},\boldsymbol{\mu}_k}=1$ for $k=x$.
The parameters $g$, $\lambda$, $a_s$, and $m$ denote the electric coupling, magnetic coupling, lattice spacing, and fermion mass, respectively. We set $a_s=1$. We initially treat $\lambda$ and $g^2$ as independent parameters and subsequently restore the standard relation
$\lambda=4/g^2$ along the physical coupling line. The plaquette operator is $\hat{P}_{\mathbf{n},\Box}=\sum_{\alpha,\beta,\gamma,\delta}
\hat{U}^{\alpha\beta}_{\mathbf{n},\boldsymbol{\mu}_x}
\hat{U}^{\beta\gamma}_{\mathbf{n}+\boldsymbol{\mu}_x,\boldsymbol{\mu}_y}
\hat{U}^{\gamma\delta\dagger}_{\mathbf{n}+\boldsymbol{\mu}_y,\boldsymbol{\mu}_x}
\hat{U}^{\delta\alpha\dagger}_{\mathbf{n},\boldsymbol{\mu}_y}$, and we use the intensive plaquette average $\langle\cos\hat B_p\rangle=\frac{1}{2N}\left\langle
\sum_{\mathbf n}\Tr \hat P_{\mathbf n,\Box}
\right\rangle$. The electric field $\hat{E}_{\mathbf{n},\boldsymbol{\mu}_k}^a$ is the associated SU$(2)$ generator, with $a$ an adjoint-representation index. Full conventions and a detailed discussion of the model are given in the companion paper~\cite{rouxinolGaugeEquivariantGaussian2026}. The Hamiltonian is invariant under local SU$(2)$ gauge transformations acting jointly on the links and fermions. Every physical state must therefore satisfy the lattice Gauss law,
$\hat{G}^a_{\mathbf n}\ket{\mathrm{phys}}=0$, a local symmetry that cannot be spontaneously broken~\cite{elitzurImpossibilitySpontaneouslyBreaking1975}.

Building on the pure-gauge SU$(2)$ construction of Ref.~\cite{spriggsAccurateGroundStates2026}, we work directly in the magnetic basis,
$\hat{U}_{\mathbf{n},\boldsymbol{\mu}_k}\ket{\mathbf U}
=
U_{\mathbf{n},\boldsymbol{\mu}_k}\ket{\mathbf U}$,
and sample link configurations from
$p(\mathbf U)=|\Psi_G(\mathbf U)|^2$.
The gauge wave function $\Psi_G(\mathbf U)$ combines a Jastrow factor with a convolutional neural network. To incorporate dynamical matter, we adapt the continuous-group Gaussian-state construction developed for the $2+1$D U$(1)$ LGT~\cite{benderVariationalMonteCarlo2023} and write
\begin{equation}
\ket{\Psi}
=
\int\mathcal D\mathbf U\,
\Psi_G(\mathbf U)
\ket{\Psi_F(\mathbf U)}
\ket{\mathbf U},
\end{equation}
where
$\mathcal D\mathbf U=\prod_{\mathbf n,k}dU_{\mathbf n,\boldsymbol{\mu}_k}$
and $\ket{\Psi_F(\mathbf U)}$ is a gauge-covariant fermionic state conditioned on $\mathbf U$.

For every sampled gauge configuration, the fermionic correction must transform covariantly under local gauge rotations so that the full many-body state remains gauge invariant and Gauss-law respecting, while retaining polynomial computational cost. We therefore choose $\ket{\Psi_F(\mathbf U)}$ to be Gaussian. Although each conditional fermionic state is Gaussian, their superposition over gauge configurations can represent non-Gaussian correlations~\cite{bravyiComplexityQuantumImpurity2017}.

At half filling, the reference state is the gauge-invariant N\'eel state $\ket{\Psi_N}$, and $\ket{\Psi_F(\mathbf U)}=\hat U_{\mathrm{corr}}(\mathbf U)\ket{\Psi_N}$ with $\hat U_{\mathrm{corr}}(\mathbf U)=\exp\!\left[i\hat{\boldsymbol{\psi}}^\dagger H_{\mathrm{full}}(\mathbf U)\hat{\boldsymbol{\psi}}\right]$, where $\hat{\boldsymbol{\psi}}$ collects all fermionic annihilation operators. 
The Hermitian generator $H_{\mathrm{full}}(\mathbf U)$ is constructed covariantly from Wilson lines and low-energy eigenvectors of the mass--hopping Hamiltonian
$\hat H_{MH}\equiv\hat H_m+\hat H_t$.
We denote its magnetic-basis matrix representation by $h_{MH}(\mathbf U)$.
In practice, $H_{\mathrm{full}}(\mathbf U)$ couples the occupied and unoccupied eigenvectors of $h_{MH}(\mathbf U)$ to each other, and is expanded in a compact set of short gauge-covariant paths, yielding $\mathcal O(L^4)$ variational parameters. 
The correction captures the competition between gauge-assisted hopping, which delocalizes the fermions, and the electric-field energy of the accompanying flux structures. 
Without it, the conditional matter state would remain the fully localized N\'eel state for every gauge configuration.

The fermionic occupation matrix defined as $[P(\mathbf U)]_{\mathbf n',\beta,\mathbf n,\alpha}=\bra{\Psi_F(\mathbf U)}\hat\psi^\dagger_{\mathbf n,\alpha}\hat\psi_{\mathbf n',\beta}\ket{\Psi_F(\mathbf U)}$, is given by $U_{\mathrm{corr}}(\mathbf U)P_N U_{\mathrm{corr}}^\dagger(\mathbf U)$, with $U_{\mathrm{corr}}(\mathbf U)=e^{iH_{\mathrm{full}}(\mathbf U)}$
and $P_N$ the N\'eel-state occupation matrix. Under local gauge rotations, both $H_{\mathrm{full}}(\mathbf U)$ and $P(\mathbf U)$ transform covariantly. Because the half-filled N\'eel reference is a local color singlet on each occupied site, $\hat U_{\mathrm{corr}}(\mathbf U)\ket{\Psi_N}$ is gauge covariant for every $\mathbf U$, and the total state satisfies Gauss's law by construction. The Gaussian structure yields analytical expressions for all fermionic contributions to the expectation value of Eq.~\eqref{eqn:HamiltonianLGT}, evaluated from $P(\mathbf U)$ and averaged over gauge configurations. The fermionic correction and gauge wave function are then optimized jointly by minimizing $\langle\hat H\rangle$.

\textbf{\textit{Results.---}}
We characterize the finite-size ground-state behavior of the $2+1$D SU$(2)$ LGT at fixed $t=1.0$ and $m=0.5$. We first treat $\lambda$ and $g^2$ independently to isolate the competition between the magnetic and mass--hopping terms. At $g^2=0$, the electric term vanishes and the gauge links form a static background. For each sampled configuration $\mathbf U$, the optimal fermionic state then follows from diagonalizing $\hat H_{MH}$, so only the gauge sector must be trained.

\begin{figure}[t!]
    \centering
    \includegraphics[width=1.0\linewidth]{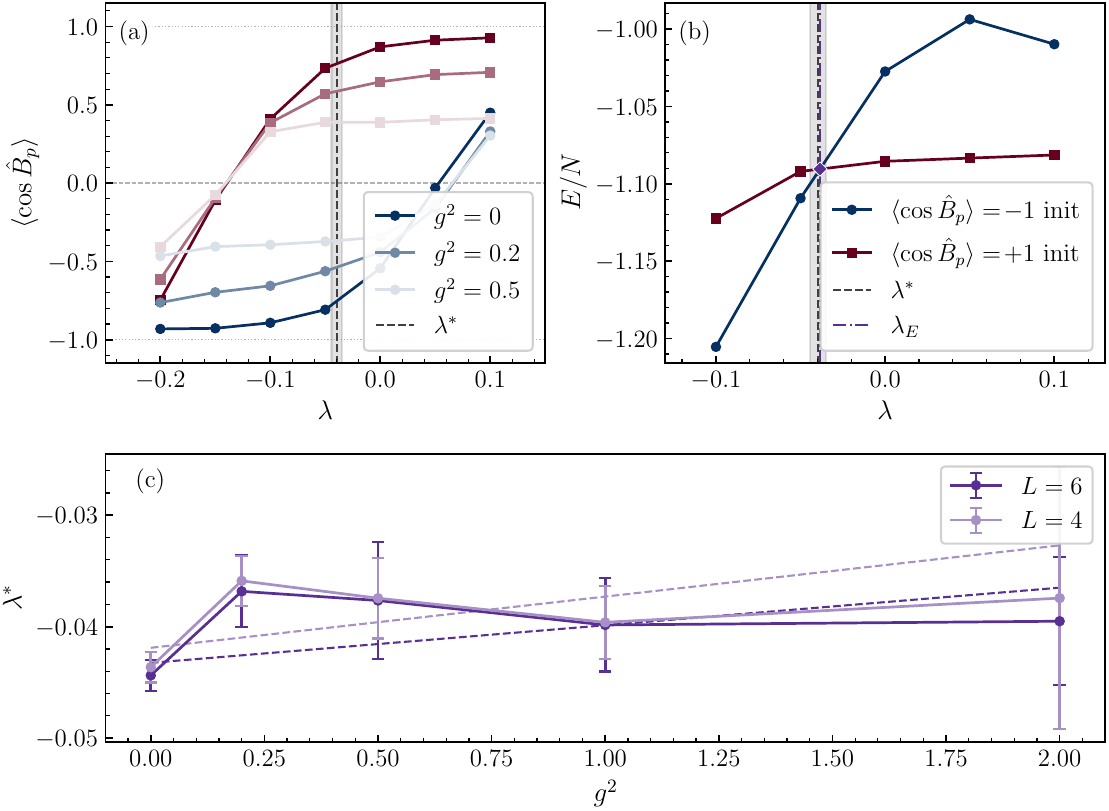}
    \caption{Magnetic-flux transition between sectors with
    $\langle\cos\hat B_p\rangle=1$ and
    $\langle\cos\hat B_p\rangle=-1$.
    (a) Hysteresis curves for $L=6$, with $\lambda^\ast=-0.040\pm 0.005$ marked by a dashed vertical line.
    (b) Energies of the two variational branches at $g^2=0$, showing an energy crossing at $\lambda_E=-0.039$ near the hysteresis estimate for $\lambda^\ast$.
    (c) Hysteresis-center estimate $\lambda^\ast$, defined as the
    location of the maximum separation between the two variational branches,
    versus $g^2$ for $L=4$ and $L=6$. All estimates agree within their
    uncertainties. The large uncertainty at $g^2=2.0$ reflects the nearly
    flat hysteresis curve. At $g^2=0$, the center of the hysteresis window is consistent with the branch-energy crossing shown in panel~(b).}
    \label{fig:fig2}
\end{figure}

To probe competing magnetic-flux sectors, we initialize two ans\"atze with $\langle\cos\hat B_p\rangle=-1$ and $\langle\cos\hat B_p\rangle=+1$ and optimize the gauge wave function across $\lambda$.
Figure~\ref{fig:fig2}(a) shows the resulting hysteresis curves for $L=6$ and $g^2=0,0.2,0.5$. The two branches interpolate between a $\pi$-flux sector, $\langle\cos\hat B_p\rangle=-1$, and a unity-flux sector, $\langle\cos\hat B_p\rangle=1$, near a small negative $\lambda$.
For $\lambda<0$, $\hat H_B$ favors the $\pi$-flux sector, whereas the mass--hopping term selects unity gauge flux. The staggering factors $\eta_{\mathbf n,\boldsymbol{\mu}_k}$ already contribute $\pi$ flux through every plaquette, so the total $\pi$ flux selected at half filling by Lieb's flux-phase theorem~\cite{liebFluxPhaseHalfFilled1994} is obtained when the gauge links contribute trivially. Because the theorem applies strictly to Abelian flux at vanishing staggered mass, we test this expectation numerically against Haar-random configurations, symmetric flux sectors, and an adversarial search in Sec.~\ref{sec:LiebCheck} of the Supplemental Material.

The coexistence of locally stable branches and the exchange of their energy ordering provide complementary evidence for a magnetic-flux transition. 
The hysteresis is a sign for the critical slowing down of convergence at a phase transition in lattice gauge studies~\cite{creutzMonteCarloStudy1979}; here, different initializations converge to distinct local minima near the transition, while Fig.~\ref{fig:fig2}(b) shows that the lower-energy branch switches in the same region for $g^2=0$ at $\lambda=\lambda_E= -0.039$. 
This is consistent with our estimate of the transition point from the maximum separation between the two branches, obtaining
$\lambda^\ast = -0.040\pm 0.005$.
The extraction procedure and uncertainties are detailed in Sec.~\ref{sec:HystCurves} of the Supplemental Material. 
An exact determination of the transition and its order would require finite-size scaling beyond the accessible system sizes.

At $g^2=0$, states with $\langle\cos\hat B_p\rangle\approx\pm1$ are associated with gauge distributions concentrated on configurations whose plaquette holonomies $P_{\mathbf n,\square}$ are predominantly close to $\pm\mathbb I$, respectively. For $g^2>0$, the electric term penalizes such sharply peaked distributions and broadens the gauge wave function, thereby flattening the hysteresis curves. Eventually, the curves are nearly flat around $\langle\cos\hat B_p\rangle\approx0$, as seen for $g^2=2$ in Sec.~\ref{sec:HystCurves} of the Supplemental Material.

From the hysteresis scans at
$g^2\in\{0,0.2,0.5,1,2\}$,
we extract $\lambda^\ast$ for $L=4$ and $L=6$.
As shown in Fig.~\ref{fig:fig2}(c), all estimates are consistent with a constant transition point within their uncertainties. Weighted linear fits yield slopes
$s_4=0.0046\pm0.0048$
and
$s_6=0.0034\pm0.0031$,
both consistent with zero at approximately one standard deviation and mutually consistent. Thus, although electric-field fluctuations progressively wash out the distinction between the two metastable branches, we resolve no corresponding drift of their crossing point. This behavior suggests that the electric term primarily broadens the gauge distribution while leaving the balance between $\hat H_{MH}$ and $\hat H_B$ nearly unchanged within the explored regime.

\begin{figure}[t!]
    \centering
    \includegraphics[width=1.0\linewidth]{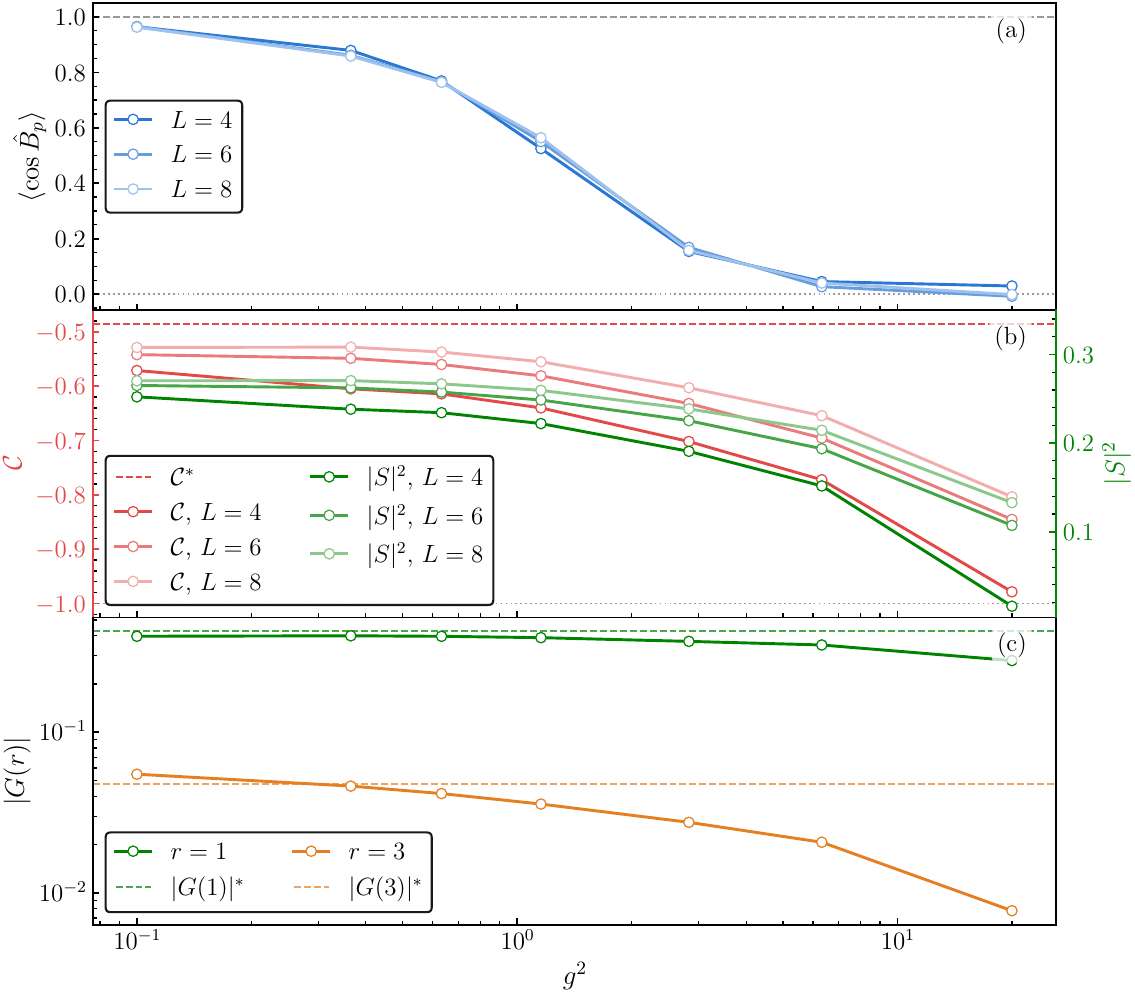}
    \caption{Ground-state observables along the physical line $\lambda=4/g^2$ for $L=4,6,8$ at $m=0.5$ and $t=1.0$.
    (a) Plaquette expectation value $\langle\cos\hat B_p\rangle$.
    (b) Chiral condensate $\mathcal C$ (red) and matter color density $|\mathbf S|^2$ (green).
    (c) Meson correlator $|G(r)|$ at $r=1,3$ for the lattice with $L=8$.
    Dashed lines indicate the $g^2\to0$ reference values obtained for
    $\hat U_{\mathbf n,\boldsymbol{\mu}_k}=\mathbb I$ in the thermodynamic limit. They agree with our results up to finite-size corrections.
    Decreasing $g^2$ drives a crossover from a flux-disordered, N\'eel-like regime to a unity-flux regime with increasingly delocalized matter and enhanced local non-singlet weight. Monte Carlo errors are smaller than the markers, hence not shown.}
    \label{fig:fig3}
\end{figure}

The negative-$\lambda$ transition lies outside the physical coupling line, for which
$\lambda=4/g^2>0$.
Along this line, the relevant competition is instead between electric-field fluctuations and the combined magnetic and mass--hopping energies. As $g^2$ decreases, the electric penalty weakens, the gauge distribution concentrates near unity-flux configurations, and gauge-assisted hopping increasingly delocalizes the fermions. Figure~\ref{fig:fig3}(a) displays the corresponding crossover from a flux-disordered regime with
$\langle\cos\hat B_p\rangle\approx0$
to a unity-flux regime with
$\langle\cos\hat B_p\rangle\approx1$.

We characterize the accompanying matter evolution through
\begin{subequations}
\begin{align}
    \mathcal{C}
    &=
    \frac{1}{N}
    \sum_{\mathbf{n}}
    (-1)^{n_x+n_y}
    \langle\hat n_{\mathbf n}\rangle,
    \label{eqn:ChiralCond}\\
    |G(r)|
    &=
    \frac{1}{2N}
    \sum_{\mathbf n,k}
    \left|\sum_{\alpha, \beta}
    \left\langle
    \hat\psi^\dagger_{\mathbf n,\alpha}
    \hat U^{\alpha\beta}_{\mathbf n,r\boldsymbol{\mu}_k}
    \hat\psi_{\mathbf n+r\boldsymbol{\mu}_k,\beta}
    \right\rangle
    \right|,
    \label{eqn:GaugeFermionCorrelator}\\
    |\mathbf S|^2
    &=
    \frac{1}{4N}
    \sum_{\mathbf n,a}
    \left\langle
    \left(
    \sum_{\alpha\beta}
    \hat\psi^\dagger_{\mathbf n,\alpha}
    \sigma^a_{\alpha\beta}
    \hat\psi_{\mathbf n,\beta}
    \right)^2
    \right\rangle,
    \label{eqn:MatterColorDensity}
\end{align}
\end{subequations}
where
$\hat U_{\mathbf n,r\boldsymbol{\mu}_k}
=
\prod_{i=0}^{r-1}
\hat U_{\mathbf n+i\boldsymbol{\mu}_k,\boldsymbol{\mu}_k}$, and $\hat{n}_\mathbf{n}=\sum_\alpha \hat{\psi}^\dagger_{\mathbf{n}, \alpha}\hat{\psi}_{\mathbf{n}, \alpha}$ the number of fermions at site $\mathbf{n}$.
All observables are evaluated as gauge-configuration averages of Wick contractions constructed from $P(\mathbf U)$. Their analytical formulas, as well as a further discussion of fermionic properties of our state are given in Sec.~\ref{sec:FermObs} of the Supplemental Material.

The chiral condensate quantifies the departure from the N\'eel reference. It approaches $\mathcal C=-1$ at strong electric coupling, where matter remains localized in the fully ordered N\'eel configuration, and its magnitude decreases as gauge-assisted hopping redistributes the fermions. This behavior is visible in Fig.~\ref{fig:fig3}(b): for every system size, $\mathcal C$ moves away from $-1$ as $g^2$ decreases.

In the opposite limit $g^2\to0$, the gauge wave function becomes concentrated near a static classical background and the fermionic problem reduces to free fermions governed by $h_{MH}(\mathbf U)$. The mass term favors occupation of the
$(-1)^{n_x+n_y}=-1$
sublattice, whereas hopping favors a spatially extended state. Their competition yields
$\lim_{g^2\to0}\mathcal C(g^2)=\mathcal C^\ast$.
Diagonalizing $h_{MH}$ at $\hat U_{\mathbf n,\boldsymbol{\mu}_k}=\mathbb I$ and taking $L\to\infty$ gives $\mathcal C^\ast=-0.486$ for $m=0.5$ and $t=1$, consistent with the trend of the variational data, up to finite-size corrections.

An independent signature is the local matter color density
$|\mathbf S|^2$, which measures the non-singlet weight in the matter sector. Gauge-assisted hopping breaks the doubly occupied on-site color singlets of the N\'eel reference and produces locally unpaired color charge connected by gauge flux. As shown in Fig.~\ref{fig:fig3}(b), $|\mathbf S|^2$ grows as $g^2$ decreases, consistently tracking the departure from the localized N\'eel regime.

The same crossover appears in the gauge-invariant meson correlator $|G(r)|$. Figure~\ref{fig:fig3}(c) shows that both $|G(1)|$ and $|G(3)|$ for $L=8$ increase as the electric coupling decreases, demonstrating the growth of gauge-dressed fermionic correlations. Their weak-coupling reference values,
$|G(1)|^\ast$ and $|G(3)|^\ast$,
are obtained from the same static unity-link problem as $\mathcal C^\ast$, and the numerical data approach these values as $g^2\to0$ in the same $L\to \infty$ limit. 
The consistent evolution of all three observables identifies a
gauge--matter delocalization crossover driven by the reduced energetic
penalty for electric-field fluctuations and the resulting enhancement
of coherent gauge-assisted hopping.

\textbf{\textit{Summary and Outlook.---}}
In this Letter, we have presented the first untruncated, variational simulation of the SU$(2)$ lattice gauge theory in $2+1$D with dynamical fermions, uncovering two distinct aspects of its finite-size ground-state physics. In the extended $(\lambda,g^2)$ parameter space, competing magnetic-flux sectors produce hysteresis and an exchange of the global energy minimum, providing evidence consistent with a phase transition near
$\lambda^\ast\approx-0.04$.
Within our numerical resolution, the transition point shows no drift with $g^2$, although increasing electric-field fluctuations progressively flatten the hysteresis curves. Along the physical coupling line
$\lambda=4/g^2$,
the system instead undergoes a gauge--matter delocalization crossover: weakening the electric coupling promotes unity flux and coherent gauge-assisted hopping, driving the matter sector away from the N\'eel reference. The chiral condensate, Wilson-line meson correlator, and local color density provide mutually consistent signatures of this evolution. Together, these results reveal how magnetic-flux structure and fermionic coherence develop across distinct regimes of an untruncated non-Abelian gauge theory with dynamical matter.

The accessible system sizes do not permit a definitive finite-size scaling analysis of the magnetic transition or the crossover. Larger lattices, more expressive gauge wave functions, and neural-network parameterizations of the matter correction could sharpen both. Varying the fermion mass would clarify how the competition between sublattice polarization and gauge-assisted hopping reshapes the ground-state behavior; in particular, the chiral limit $m=0$ provides a natural setting in which to investigate spontaneous chiral-symmetry breaking. Denser coupling scans would further enable susceptibility-based characterizations of the crossovers.

More broadly, the continuous-group Hamiltonian formulation provides direct access to regimes where Euclidean Monte Carlo is obstructed by the sign problem, including finite chemical potential and topological $\theta$ terms. Extensions to thermal and excited states and to real-time dynamics could address string breaking, thermalization, and nonequilibrium gauge--matter transport in non-Abelian theories~\cite{gonzalez-cuadraObservationStringBreaking2025,xuStringBreakingDynamics2025,cataldiRealTimeStringDynamics2025}. The present results therefore establish a route toward first-principles studies of nonperturbative non-Abelian gauge--matter physics beyond the equilibrium regimes accessible to conventional Euclidean methods.

\bigskip
\footnotesize
\begin{acknowledgments}
    \textbf{\textit{Acknowledgments.---}} We are grateful to Jannes Nys and Thomas Spriggs for stimulating discussions.
    G.R. and J.C.H.~acknowledge funding by the Max Planck Society, the Deutsche Forschungsgemeinschaft (DFG, German Research Foundation) under Germany's Excellence Strategy - EXC-2111 - 390814868, and the European Research Council (ERC) under the European Union's Horizon Europe research and innovation program (Grant Agreement No.~101165667)-ERC Starting Grant QuSiGauge.
    This work is part of the Quantum Computing for High-Energy Physics (QC4HEP) working group.
    J.B. is supported by a Feodor Lynen Research Fellowship from the Alexander von Humboldt Foundation.
    P.E. acknowledges the support received from the Dutch National Growth Fund (NGF) as part of the Quantum Delta NL program in the NWO-Quantum Technology program (Grant No.~NGF.1623.23.006).
    P.E. also acknowledges funding from the Carl-Zeiss-Stiftung (CZS Center QPhoton).
    M.G. is supported by CERN through the CERN Quantum Technology Initiative. M.G. thanks ESA SpaceHPC for the time provided on their infrastructure.
\end{acknowledgments}

\textbf{\textit{Data Availability.---}}
The data and the code used to generate all plots in this work are available in~\cite{rouxinolDataQuantumPhase2026}.
\normalsize

\bibliographystyle{apsrev4-2}
\bibliography{biblio}
\clearpage
\onecolumngrid

\setcounter{equation}{0}
\setcounter{figure}{0}
\setcounter{table}{0}
\setcounter{supplem}{0}
\setcounter{page}{1}

\renewcommand{\theequation}{S\arabic{equation}}
\renewcommand{\thefigure}{S\arabic{figure}}
\renewcommand{\thetable}{S\arabic{table}}
\renewcommand{\theHequation}{S\arabic{equation}}
\renewcommand{\theHfigure}{S\arabic{figure}}
\renewcommand{\theHtable}{S\arabic{table}}

\begin{center}
    {\Large \textbf{Supplemental Material for ``Quantum Phase Diagram of the $2+1$D Untruncated SU$(2)$ Lattice Gauge Theory with Dynamical Fermions''}}\\
    Gabriel Rouxinol, Julian Bender, Patrick Emonts, Michele Grossi, and Jad C.~Halimeh
\end{center}

\twocolumngrid

\section{Numerical test of the Lieb flux expectation}
\refstepcounter{supplem}\label{sec:LiebCheck}

Because the staggering already supplies $\pi$ flux per plaquette, Lieb's flux-phase theorem~\cite{liebFluxPhaseHalfFilled1994} leads us to expect $\hat{U}_{\mathbf{n},\boldsymbol{\mu}_k}=\mathbb{I}$ to minimize the mass--hopping energy. As the theorem was proved only for Abelian flux on bipartite lattices satisfying certain conditions, we tested this expectation numerically. We constructed states within the same flux sector, meaning with the same $\langle\cos \hat B_p\rangle$, compared $8000$ Haar-random link configurations spanning $L\in\{4,6,8\}$, and scanned a family of gauge configurations defined by continuous parameters that interpolate between different values of $\langle\cos \hat B_p\rangle$. This was done for mass-to-hopping ratios $\frac{m}{t}\in\{0.1,0.5,1.0,2.0\}$. For $L\geq6$ we find no configuration whose energy lies more than $0.2\%$ below that of $\hat U=\mathbb I$. At $L=4$ and for the parameters used throughout this work ($m/t=0.5$), around $5.4\%$ of the tested configurations had a lower energy than $\hat{U}_{\mathbf{n},\boldsymbol{\mu}_k}=\mathbb{I}$. However, the difference was at most $1\%$ and occurred for configurations with $\langle\cos \hat B_p\rangle\approx0.8-1$, showing that the competition between the magnetic and fermionic terms still favors the unity-flux sector. The percentage of configurations with lower energy and their relative improvement decrease with system size $L$. This is consistent with the flux-counting expectation becoming exact in the thermodynamic limit. Furthermore, this violation is more pronounced at smaller $\frac{m}{t}$, highlighting how the hopping Hamiltonian is responsible for the finite-size effects on small lattices seen here. The total Hamiltonian therefore contains two terms that favor opposite extreme values of $\cos \hat B_p$ when $\lambda<0$. In that regime, we expect one region with $\langle\cos \hat B_p\rangle\approx-1$ and another with $\langle\cos \hat B_p\rangle\approx1$, producing a transition between the two regimes.

\section{Hysteresis curve for the flux transition}
\refstepcounter{supplem}\label{sec:HystCurves}

\begin{figure}[ht]
    \centering
    \includegraphics[width=1.0\linewidth]{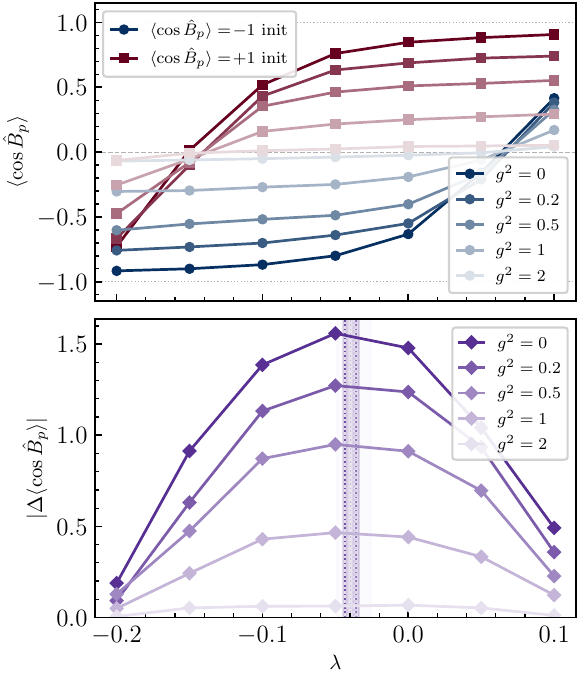}
    \caption{Hysteresis scan of $\langle\cos \hat B_p\rangle$ versus $\lambda$ for $g^2\in\{0,0.2,0.5,1,2\}$ and $L=4$. The upper panel shows the branches initialized at $\langle\cos\hat B_p\rangle=-1$ (blue) and $\langle\cos\hat B_p\rangle=+1$ (red); the lower panel shows their separation $|\Delta\langle\cos\hat B_p\rangle|$. Increasing $g^2$ progressively flattens the curves and reduces their separation. Vertical lines indicate the estimated $\lambda^\ast$ and its uncertainty. The fitted maxima remain near $\lambda^\ast\approx-0.04$.}
    \label{fig:AllgHysterisis}
\end{figure}

\begin{figure}[ht]
    \centering
    \includegraphics[width=1.0\linewidth]{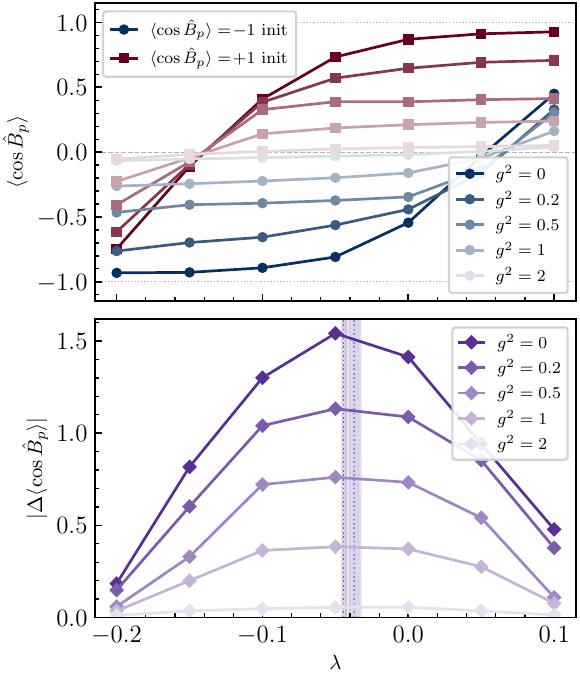}
    \caption{Same as Fig.~\ref{fig:AllgHysterisis}, but for $L=6$. The estimated transition points again remain near $\lambda^\ast\approx-0.04$.}
    \label{fig:AllgHysterisis_L6}
\end{figure}

To study the transition between the two magnetic phases when $(g^2,\lambda)$ are treated as independent parameters, we initialize the system at $\langle\cos \hat B_p\rangle=\pm1$ and train the model at $g^2=0$, then use the resulting parameters to initialize the calculations at $g^2\neq0$. We obtain the final $\langle\cos \hat B_p\rangle$ for both initializations throughout the $(g^2,\lambda)$ parameter space. The corresponding curves are shown for $L=4$ in Fig.~\ref{fig:AllgHysterisis} and for $L=6$ in Fig.~\ref{fig:AllgHysterisis_L6}. In each figure, the upper panel displays the two branches and the lower panel their absolute separation. To estimate $\lambda^\ast$, we select the five points closest to the maximum branch separation and fit them with a quadratic polynomial. The quoted uncertainty is obtained from this fit and is comparatively large because only a few points constrain the maximum. At large $g^2$, the hysteresis curves flatten and the estimate of $\lambda^\ast$ consequently becomes less precise. We take as an estimate of $\lambda^\ast$ the average over all used $g^2$ values and the uncertainty as the highest deviation from the mean, obtaining for $L=4$ $\lambda^\ast=-0.039\pm0.005$ and for $L=6$ $\lambda^\ast=-0.040\pm 0.005$. 

\section{Fermionic observables}
\begin{figure}[t!]
    \centering
    \includegraphics[width=1.0\linewidth]{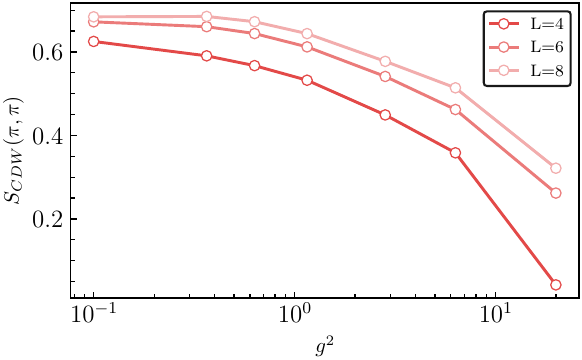}
    \caption{Evolution of the charge--density wave structure factor at momentum $\mathbf{k}=(\pi,\pi)$, $\langle S_{CDW}(\pi,\pi)\rangle$ as a function of $g^2$ and for lattice sizes $L=4,6,8$. Its increase as the electric coupling decreases is another sign of a delocalized fermionic state.}
    \label{fig:SCDW}
\end{figure}

\refstepcounter{supplem}\label{sec:FermObs}
As discussed in the main text, one of the main advantages of the Gaussian formulation is that computing fermionic observables reduces to products between matrices and the fermionic occupation matrix $P(\mathbf{U})$, traced and averaged over Monte Carlo samples. Here $P(\mathbf{U})$ and the vector $\hat{\boldsymbol{\psi}}$ collecting all fermionic annihilation operators are those defined in the main text. Using properties of Gaussian states, one finds that for any one-body fermionic operator that is diagonal in the magnetic basis, $\hat{O}=\int \mathcal{D}\mathbf{U}\,\hat{\boldsymbol{\psi}}^\dagger O(\mathbf U)\hat{\boldsymbol{\psi}}\ket{\mathbf{U}}\bra{\mathbf{U}}$, the expectation value follows
\begin{equation}
    \langle\hat{O}\rangle =\int \mathcal{D}\mathbf{U}p(\mathbf{U})\Tr(P(\mathbf{U})O(\mathbf{U})),
\end{equation}
while higher-order operators can be extracted using Wick contractions. In the simple case where the matrix $O(\mathbf{U})=\mathbb{I}$, we obtain $\langle\hat{\boldsymbol{\psi}}^\dagger \hat{\boldsymbol{\psi}}\rangle=\sum_\mathbf{n}\langle\hat{n}_\mathbf{n}\rangle=N_f$, the total fermion number, which is conserved and equals $N$ at half filling. We define $P_{\mathbf{n}}(\mathbf{U})\equiv P(\mathbf{U})_{\mathbf{n}\mathbf{n}}$ for the $2\times2$ on-site color block of $P(\mathbf{U})$, so that $\Tr P_{\mathbf{n}}(\mathbf{U})=\langle\hat{n}_\mathbf{n}\rangle$, and $P_{\mathbf{n}\mathbf{n}'}(\mathbf{U})$ for the corresponding off-diagonal site block. Using what we described above, the chiral condensate of a single sample has the analytical formula 
\begin{equation}
	\mathcal{C}(\mathbf{U})=\frac{1}{N}\sum_{\mathbf{n}}(-1)^{n_x+n_y}\,
	\Tr P_{\mathbf{n}}(\mathbf{U}),
	\label{eqn:ChiralCondPerConfig}
\end{equation}
which is then averaged over all sampled configurations. Since the staggered sign appearing in Eq.~\eqref{eqn:ChiralCondPerConfig} is the same one entering $\hat{H}_m$, the condensate requires no separate evaluation: it is fixed by the mass contribution to the energy through $\mathcal{C}=\langle\hat{H}_m\rangle/(mN)$. To compute the expectation value $\left\langle\hat\psi^\dagger_{\mathbf n,\alpha}\hat U^{\alpha\beta}_{\mathbf n,r\boldsymbol{\mu}_k}\hat\psi_{\mathbf n+r\boldsymbol{\mu}_k,\beta}\right\rangle$, one first needs the matrix representation of $\hat{U}^{\alpha\beta}_{\mathbf n,r\boldsymbol{\mu}_k}$, which is simply the path-ordered product $U_{\mathbf{n},r\boldsymbol{\mu}_k}=\prod_{i=0}^{r-1}U_{\mathbf{n}+i\boldsymbol{\mu}_k,\boldsymbol{\mu}_k}$, as it is a diagonal operator in the magnetic basis. Using the formulation above, we obtain that the expectation value for a single sample, denoted by $\langle\cdot\rangle_\mathbf{U}$, is
\begin{equation}
	\sum_{\alpha\beta}\Big\langle \hat{\psi}^{\dagger}_{\mathbf{n},\alpha}\,
	U^{\alpha\beta}_{\mathbf{n},r\boldsymbol{\mu}_k}\,
	\hat{\psi}_{\mathbf{n}+r\boldsymbol{\mu}_k,\beta}\Big\rangle_{\mathbf{U}}
	=\Tr\!\left[U_{\mathbf{n},r\boldsymbol{\mu}_k}\,
	P_{\mathbf{n}+r\boldsymbol{\mu}_k,\,\mathbf{n}}(\mathbf{U})\right].
	\label{eqn:MesonPerConfig}
\end{equation}
Each such trace is separately gauge invariant, since $U_{\mathbf{n},r\boldsymbol{\mu}_k}\to\Omega_{\mathbf{n}}U_{\mathbf{n},r\boldsymbol{\mu}_k}\Omega^\dagger_{\mathbf{n}+r\boldsymbol{\mu}_k}$ compensates the covariant transformation of the occupation-matrix block. 
This also allows us to compute the expectation value of squared operators such as the matter color density $|\mathbf{S}|^{2}$ of the main text. Introducing the on-site color operator $\hat{S}^a_{\mathbf{n}}=\sum_{\alpha\beta}\hat{\psi}^{\dagger}_{\mathbf{n},\alpha}\sigma^a_{\alpha\beta}\hat{\psi}_{\mathbf{n},\beta}$, it reads
\begin{equation}
    |\mathbf{S}|^{2}=\frac{1}{4N}\sum_{\mathbf{n}}\left\langle \sum_a\big(\hat{S}^a_{\mathbf{n}}\big)^2\right\rangle .
    \label{eqn:ColorMatterDensity}
\end{equation}
Using the Fierz identity $\sum_a\sigma^a_{\alpha\beta}\sigma^a_{\gamma\delta}=2\delta_{\alpha\delta}\delta_{\beta\gamma}-\delta_{\alpha\beta}\delta_{\gamma\delta}$, the on-site color Casimir obeys the operator identity
\begin{equation}
	\hat{S}^2_{\mathbf{n}}=\sum_a\big(\hat{S}^a_{\mathbf{n}}\big)^2 = 3\,\hat{n}_{\mathbf{n}}\big(2-\hat{n}_{\mathbf{n}}\big),
\end{equation}
which vanishes at $\hat{n}_{\mathbf{n}}=0,2$ (empty site, or the color singlet) and equals its maximum value of $3$ at $\hat{n}_{\mathbf{n}}=1$, as expected for a single color charge. 
Taking the expectation value and using the Wick contractions to obtain $\langle\hat{n}_{\mathbf{n}}^2\rangle=\big(\Tr P_{\mathbf{n}}(\mathbf{U})\big)^2+\Tr P_{\mathbf{n}}(\mathbf{U})-\Tr\big(P_{\mathbf{n}}(\mathbf{U})^2\big)$, we get
\begin{equation}
	\left\langle\hat{S}^2_{\mathbf{n}}\right\rangle =3\bigg[ \Tr\bigg(P_{\mathbf{n}}(\mathbf{U})\big(\mathbb{I}+P_{\mathbf{n}}(\mathbf{U})\big)\bigg) - \big(\Tr P_{\mathbf{n}}(\mathbf{U})\big)^2\bigg],
\end{equation}
again averaged as $\langle|\mathbf{S}|^2\rangle=\int\mathcal{D}\mathbf{U}\,p(\mathbf{U})\,|\mathbf{S}|^2(\mathbf{U})$. Another second-order correlator not considered in the main text is the connected density-density correlator $C(\mathbf{n},\mathbf{n}')=\langle \hat{n}_\mathbf{n} \hat{n}_{\mathbf{n}'}\rangle-\langle\hat{n}_{\mathbf{n}}\rangle\langle\hat{n}_{\mathbf{n}'}\rangle$, which has the equivalent form
\begin{equation}
	C(\mathbf{n},\mathbf{n}') = \delta_{\mathbf{n}\mathbf{n}'}\Tr P_{\mathbf{n}}(\mathbf{U}) - \Tr\left[P_{\mathbf{n}\mathbf{n}'}(\mathbf{U})P_{\mathbf{n}'\mathbf{n}}(\mathbf{U})\right],
	\label{eqn:Cnnprime}
\end{equation}
so that computing the charge--density wave structure factor at momentum $\mathbf{k}=(\pi,\pi)$, $\langle S_{CDW}(\pi,\pi)\rangle$, defined as 
\begin{equation}
	S_{CDW}(\pi,\pi)(\mathbf{U}) = \frac{1}{N}\sum_{\mathbf{n},\mathbf{n}'} (-1)^{n_x+n_y+n_x'+n_y'}C(\mathbf{n},\mathbf{n}'),
    \label{eqn:CDWfactor}
\end{equation}
also becomes a Monte Carlo average over traces of the occupation matrix. Because $C(\mathbf{n},\mathbf{n}')$ is connected, $\langle S_{CDW}(\pi,\pi)\rangle$ equals the variance per site of the staggered density, and therefore probes fluctuations around the mean staggered order rather than the order parameter itself, capturing non-local correlations to which the chiral condensate is insensitive. As for the observables in the main text, it increases as $g^2$ decreases, as seen in Fig.~\ref{fig:SCDW}, again reinforcing that the fermionic state becomes more delocalized as the electric coupling decreases.

\end{document}